\documentclass[superscriptaddress,twocolumn,10pt,showpacs,aps,prl,longbibliography]{revtex4-1}
\usepackage{amsmath}
\usepackage{graphicx}
\usepackage{inputenc}
\usepackage{color}
\usepackage{setspace}

\definecolor{green}{rgb}{0.46,0.93,0}

\begin{document}
\newcommand{\rt}{R$_2$T$_2$O$_7$}
\newcommand{\przr}{Pr$_2$Zr$_2$O$_7$}
\newcommand{\tbhf}{Tb$_2$Hf$_2$O$_7$}
\newcommand{\prsn}{Pr$_2$Sn$_2$O$_7$}
\newcommand{\prir}{Pr$_2$Ir$_2$O$_7$}
\newcommand{\prhf}{Pr$_2$Hf$_2$O$_7$}
\newcommand{\prru}{Pr$_2$Ru$_2$O$_7$}
\newcommand{\ndzr}{Nd$_2$Zr$_2$O$_7$}
\newcommand{\hoti}{Ho$_2$Ti$_2$O$_7$}
\newcommand{\tbti}{Tb$_2$Ti$_2$O$_7$}
\newcommand{\tbsn}{Tb$_2$Sn$_2$O$_7$}
\newcommand{\erti}{Er$_2$Ti$_2$O$_7$}
\newcommand{\ybti}{Yb$_2$Ti$_2$O$_7$}
\newcommand{\dyti}{Dy$_2$Ti$_2$O$_7$}
\newcommand{\biti}{Bi$_2$Ti$_2$O$_7$}
\newcommand{\lazr}{La$_2$Zr$_2$O$_7$}
\newcommand{\nd}{Nd$^{3+}$}
\newcommand{\pr}{Pr$^{3+}$}
\newcommand{\ho}{Ho$^{3+}$}
\newcommand{\dy}{Dy$^{3+}$}
\newcommand{\er}{Er$^{3+}$}
\newcommand{\tb}{Tb$^{3+}$}
\newcommand{\zr}{Zr$^{4+}$}

\newcommand{\mub}{$\mu_B$}
\title{Disorder and Quantum spin ice}
\author{N. Martin}
\affiliation{Laboratoire L\'eon Brillouin, CEA, CNRS, Universit\'e Paris-Saclay, CEA-Saclay, F-91191 Gif-sur-Yvette, France}
\author{P. Bonville}
\affiliation{SPEC, CEA, CNRS, Universit\'e Paris-Saclay, CEA-Saclay, F-91191 Gif-sur-Yvette, France}
\author{E. Lhotel}
\affiliation{Institut N\'eel, CNRS and Univ. Grenoble Alpes, F-38042 Grenoble, France}
\author{S. Guitteny}
\affiliation{Laboratoire L\'eon Brillouin, CEA, CNRS, Universit\'e Paris-Saclay, CEA-Saclay, F-91191 Gif-sur-Yvette, France}
\author{A. Wildes}
\affiliation{Institut Laue Langevin, 38042 Grenoble, France}
\author{C. Decorse}
\affiliation{ICMMO, Universit\'e Paris-Sud, Universit\'e Paris-Saclay, F-91405 Orsay, France}
\author{M. Ciomaga Hatnean}
\affiliation{Department of Physics, University of Warwick, Coventry, CV4 7AL, United Kingdom}
\author{G. Balakrishnan}
\affiliation{Department of Physics, University of Warwick, Coventry, CV4 7AL, United Kingdom}
\author{I. Mirebeau}
\affiliation{Laboratoire L\'eon Brillouin, CEA, CNRS, Universit\'e Paris-Saclay, CEA-Saclay, F-91191 Gif-sur-Yvette, France}
\author{S. Petit}
\email[]{sylvain.petit@cea.fr}
\affiliation{Laboratoire L\'eon Brillouin, CEA, CNRS, Universit\'e Paris-Saclay, CEA-Saclay, F-91191 Gif-sur-Yvette, France}
\date{\today}

\begin{abstract}
We report on diffuse neutron scattering experiments providing evidence for the presence of random strains in the quantum spin ice candidate \przr. Since \pr\ is a non-Kramers ion, the strain deeply modifies the picture of Ising magnetic moments governing the low temperature properties of this material. It is shown that the derived strain distribution accounts for the temperature dependence of the specific heat and of the spin excitation spectra. Taking advantage of mean field and spin dynamics simulations, we argue that the randomness in \przr\, promotes a new state of matter, which is disordered, yet characterized by short range antiferroquadrupolar correlations, and from which emerge spin-ice like excitations. This study thus opens an original research route in the field of quantum spin ice.
\end{abstract}


\maketitle 


In condensed matter physics, disorder usually tends to freeze the degrees of freedom. This is for instance the case in the Anderson localization phenomenon \cite{Anderson1958} or in the spin glass transition \cite{Edwards1975}. In frustrated systems, disorder is also expected to dramatically impact the ground state, possibly engendering new states of matter. For instance, in pyrochlore antiferromagnets, considered as the archetype of 3D geometrically frustrated systems, weak structural and magnetic disorder can induce spin-glass \cite{Saunders2007,Andreanov2010} or even ``topological spin glass" phases \cite{Sen2015} in spin ice. When the magnetic moment in the latter materials is formed by non-Kramers ions, intrinsically very sensitivity to strain, disorder or local distortions, as for instance in \tbti, \tbhf, \przr\ or \prsn, the physics is even more intriguing. A bunch of experimental studies reported the lack of long range order down to very low temperature along with an unexpected spin fluctuation spectrum, indicating a strong density of low energy excitations \cite{Gingras14,Bonville2011,Petit12-tbsn,Foronda15,Duijn05,Kimura13,Wen2017,Sibille16}. It was then suggested that instead of driving glassy behavior, disorder may open a new route in stabilizing the enigmatic ``quantum spin ice'' state \cite{Savary2017,Wen2017}. 

\begin{figure}[!t]
\center{\includegraphics[width=7.5cm]{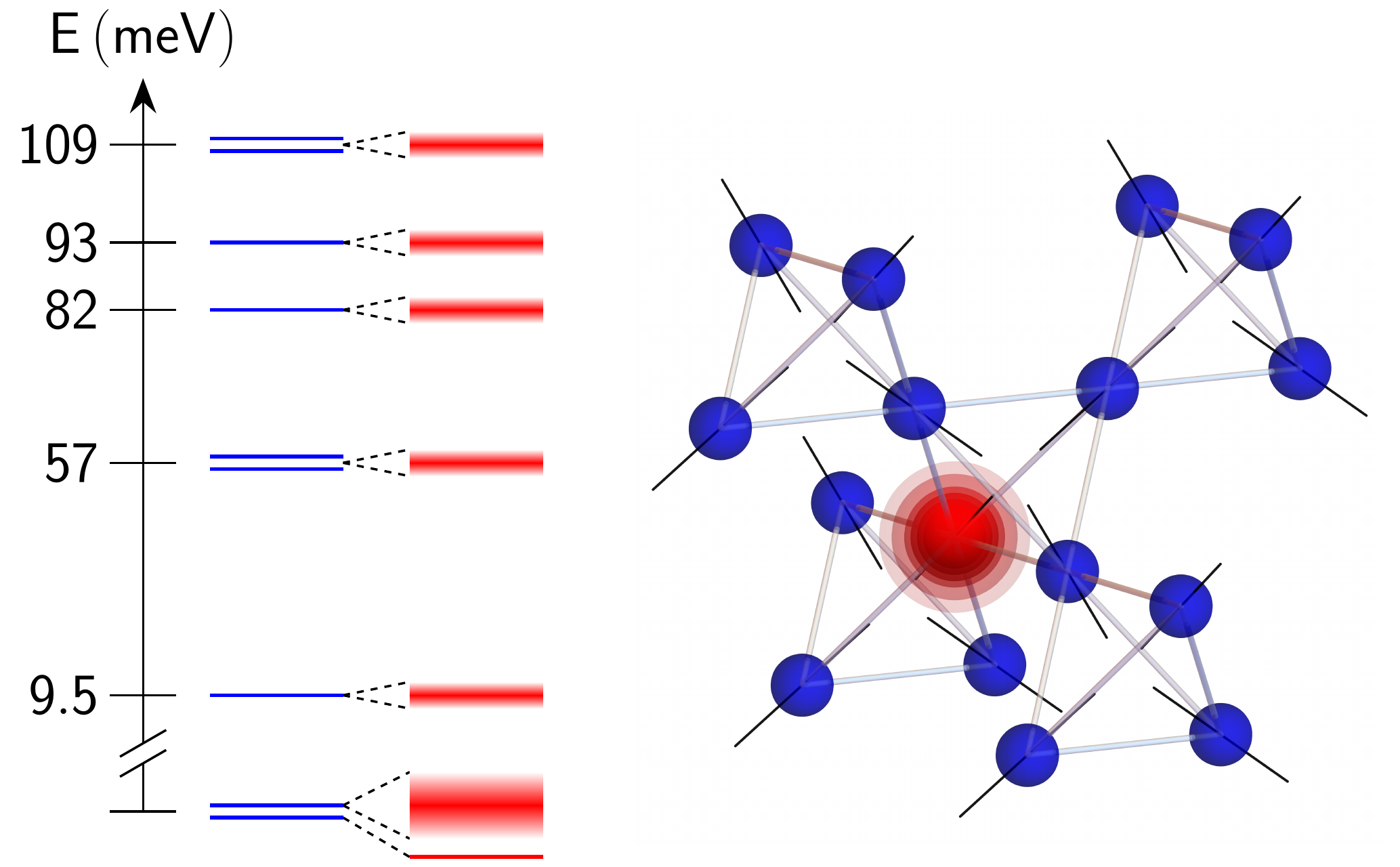}}
\caption{\label{dessin} {\bf Defects in \przr.} The strain induced by a defect in the pyrochlore structure is sketched using red concentric spheres. The CEF scheme \cite{Kimura13,Petit16} is shown schematically on the left side, for the unperturbed (blue) and perturbed (red) sites. Double (resp. single) blue lines correspond to doublets (singlets). The shaded red rectangles illustrate the broadening of the CEF level by disorder. Finally, the black lines indicate the local CEF axis directions.}
\end{figure}

While this scenario appears extremely appealing, raising eventually the hope of tuning new exotic states of matter by disorder, without even touching the magnetic lattice, it remains to show that the \emph{actual} disorder in real materials is indeed consistent with these ideas. In this work, we address these issues in the particular case of \przr\ which is ideally documented among quantum spin ice candidates. 
We first report on neutron scattering measurements providing evidence for an intrinsic distribution of strains, ensuring a direct experimental foundation for our point. We then examine in details the consequences of this randomness on the low temperature magnetic properties. It is found that this random strain distribution indeed accounts for a number of experimental features. Eventually, we find that it corresponds to a strong perturbation. \przr\, then hosts a competition between disorder and the ``native'' effective interactions, namely the Ising coupling ${\cal J}^{zz}$ (considered in Ref \onlinecite{Savary2017}) and the transverse quadrupolar terms ${\cal J}^{\pm}$, known to play a prominent role in \przr\ \cite{Petit16}. With the support of real space mean field and time-evolving spin dynamics simulations, we show that the competition with ${\cal J}^{\pm}$ especially promotes a new kind of spin liquid, disordered (or weakly ordered), characterized by short-range antiferroquadrupolar correlations and from which emerge a peculiar spin-ice like excitation spectrum. 

This study takes advantage of an abundant literature on \przr. As sketched in Figure \ref{dessin}, the physics is governed by the $\uparrow$ and $\downarrow$ degenerate states of the rare earth crystal field (CEF) ground state doublet. The corresponding magnetic moments sit on the vertices of the pyrochlore lattice (made of corner sharing tetrahedra) and point along local $\langle 111 \rangle$ directions (see Figure \ref{dessin}). In \przr, the $\left\{\left|\uparrow\rangle\right.,\left|\downarrow\rangle\right.\right\}$ subspace is well protected since the first CEF excited state is located at $\simeq$ 10 meV \cite{Matsuhira09, Kimura13, Monica14, Sibille16}. Furthermore, the Curie-Weiss temperature inferred from magnetic susceptibility is negative, indicating antiferromagnetic interactions. The specific heat shows a broad peak at about 2 K \cite{Lutique04,Matsuhira09, Kimura13, Sibille16, Nakatsuji06, Zhou08}, bearing some resemblance with the classical spin ice \dyti,\ along with an upturn at low temperature attributed to the hyperfine contribution. Above 1 T, the broad anomaly shifts to larger temperatures \cite{Petit16}. Finally, the magnetic excitation spectrum $S({\bf Q},\omega)$, measured by inelastic neutron scattering, has been described as a dynamical spin ice mode consisting of a broad inelastic response centered around $\Delta \approx$ 0.4 meV. Its structure factor strongly resembles the famous spin-ice pattern, with pinch points and arm-like features along $\langle hhh \rangle$ directions \cite{Kimura13,Petit16,Wen2017}. 

In the particular case of non-Kramers ions, the degeneracy of the ground doublet is easily lifted by perturbations, defects, local deformations and strains which, by virtue of the magneto-elastic interaction, perturb the electronic density over the rare earth sites. The Ising $\left| \uparrow \downarrow \rangle\right.$ doublet turns into ``tunnel-like'' wave-functions \cite{Petit12-tbsn,Foronda15,Duijn05,Wen2017}
\begin{equation} \label{tunn}
\vert a \rangle \simeq \frac{1}{\sqrt{2}}\ [\left|\uparrow\rangle\right. - \left|\downarrow\rangle\right.]~
~ {\rm and}\ \ \vert s \rangle \simeq \frac{1}{\sqrt{2}}\ [\left|\uparrow\rangle\right. + \left|\downarrow\rangle\right.] \quad ,
\end{equation}
and split by a random quantity $\Delta$ which directly reflects the strength of the perturbations (see Figure \ref{dessin}). In this local picture, as we show below, the specific heat consists in multiple Schottky anomalies due to the distribution of splittings and the large width of the spin excitations spectrum arises from the transitions within the randomly split doublets. These transitions can be observed by neutrons because their cross section, proportional to $\langle a \vert {\bf J}_z \vert s \rangle = \langle \uparrow \vert {\bf J}_z \vert \uparrow \rangle$, is non-zero. This is at variance with the unperturbed case since the non-Kramers character imposes $|\langle \uparrow | {\bf J} | \downarrow \rangle| \equiv 0$. 


\section{Results}
\subsection{Lattice strain and diffuse scattering}

\begin{figure*} [t]
\center{\includegraphics[width=\textwidth]{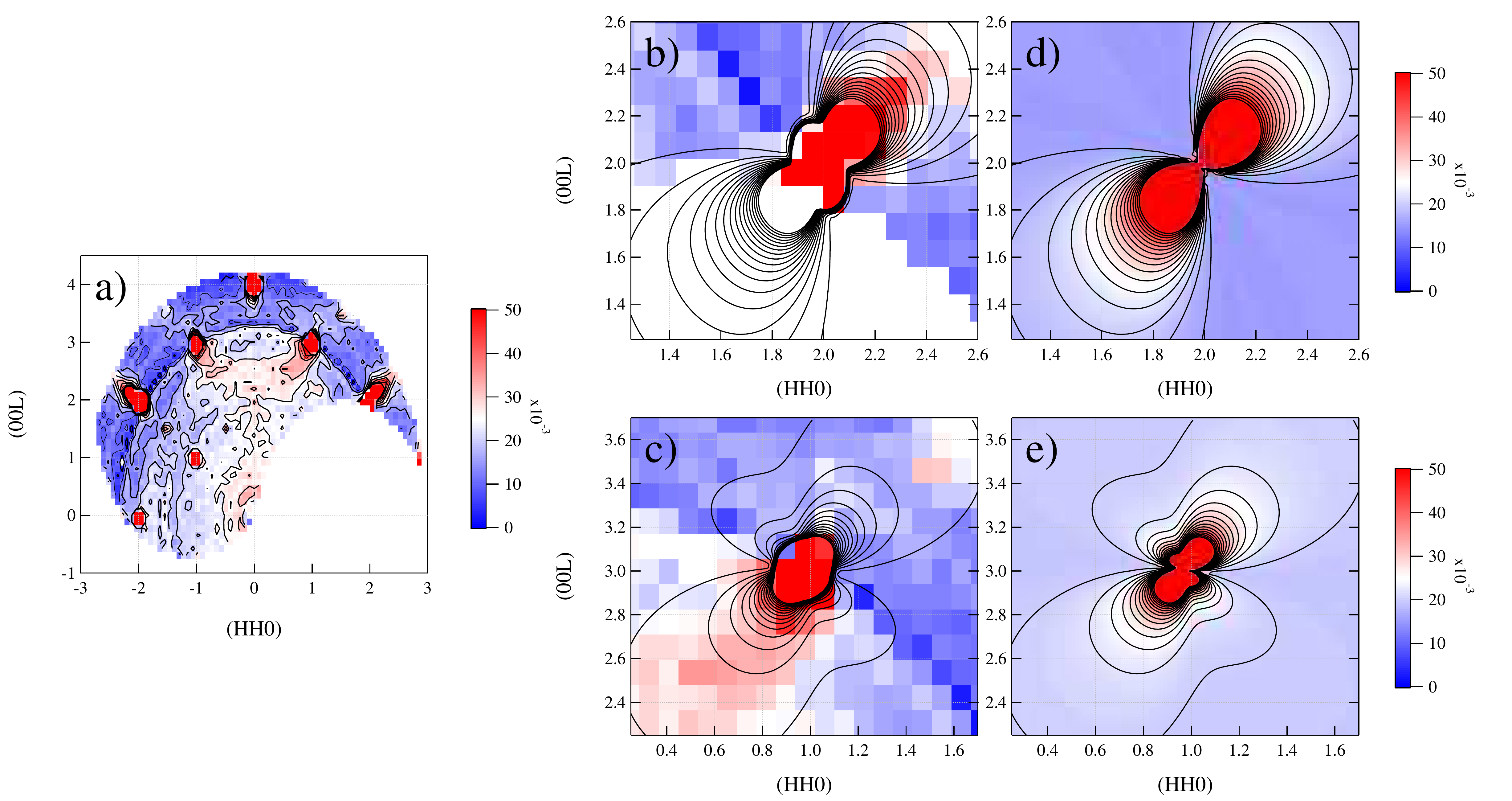}}
\caption{\label{DiffuseNSF} {\bf Elastic scattering on \przr.} {\bf (a)} Diffuse (Non Spin Flip) scattering map measured at 50\, mK in \przr. {\bf (b,c)} Zoom of the experimental data around the (113) and (222) Bragg peaks. {\bf (d,e)} Fit to the data of the diffuse scattering model discussed in the text. The black lines in {\bf (b,c)} show the calculated iso-intensity lines for the total scattering, while in {\bf (d,e)} they correspond to the Huang scattering only. Intensities are in arbitrary units.}
\end{figure*}

\przr\, samples have been prepared as carefully as possible, and especially annealed to prevent the presence of Pr$^{4+}$ ions, resulting in a beautiful green transparent color \cite{Monica15}. Their crystalline structure is very well refined by a perfect pyrochlore structure, the amount of defects being beyond the accuracy of standard diffraction, typically 1\% \cite{Monica14,Petit16}. The pyrochlore structure of \przr\ actually contains two types of cations, \pr\, and \zr\, on 16d and 16c sites respectively. The two oxygen sites O1 (48f) and O2 (8b) are fully occupied, whereas the 8a site is vacant. 

Lattice strains in \przr\, are first evidenced by means of polarized neutron scattering measurements. Fig. \ref{DiffuseNSF} shows Non-Spin-Flip (NSF) diffuse scattering maps recorded in the (HH0, 00L) scattering plane (see Supplemental Information). Anomalies appear in the vicinity of the nuclear Bragg reflections (222), (311) and (400), taking the form of butterfly features elongated along specific directions (Fig. \ref{DiffuseNSF}). In contrast, the nuclear Bragg peaks (111) and (220) show quasi isotropic shape with negligible diffuse scattering. The measurements were repeated at three temperatures (10\,K, 1\,K and 50\,mK). No significant evolution could be noticed, hence suggesting that this diffuse scattering is associated with structural disorder and lattice strains.
 
In many ordered pyrochlores, the less energetic defects mechanisms \cite{Minervini2000} are: (i) cation antisite disorder, namely substitution of R$^{3+}$ cations by T$^{4+}$ ones or the opposite; (ii) creation of anion Frenkel pair, consisting of a vacant $\it{48f}$ site and an interstitial oxygen occupying a $\it{8a}$ site. The association of both processes is often at play. It influences electronic or ionic conductivity, and helps the accommodation of radiation damages and the resistance to amorphization \cite{Sickafus2007}. However, these disordered fluorite-like defects are not the only source of lattice strains in the ordered pyrochlore structure. Other mechanisms involving lattice displacements could be at play. For example, tiny or short range ordered distortions lead to dimer-like magnetic excitations in GeCo$_2$O$_4$ ordered spinel \cite{Tomiyasu2011} as well as a kind of frustrated ferroelectricity in niobium based pyrochlores \cite{McQueen2008}. In \przr, a local off-centering of the \pr\ ions from the ideal pyrochlore $\it{16d}$ sites has been reported \cite{Koohpayeh2014} as the dominant mechanism, similar to observations for the trivalent cation in \biti\ \cite{Shoemaker2011} and \lazr\ \cite{Tabira1999}.


Anyhow, irrespective of the actual nature of disorder, the lattice strains can be analyzed using the same formalism. The diffuse scattering maps have been worked out in the framework of the so-called Huang scattering \cite{Huang1947} (see Supplemental Information), assuming random static displacements of the atoms of the structure. The scattered intensity at a reciprocal lattice vector ${\bf Q}={\bf \tau}+{\bf q}$ around a Bragg peak ${\bf \tau}$ is written:
\begin{equation}
	I\left({\bf Q}\right) = \left|F\left({\bf \tau}\right)\right|^{2}\cdot\left\{\mathcal{S}_{\rm Bragg}\left({\bf \tau}\right)+ c \, \mathcal{S}_{\rm Diff}\left({\bf Q}\right)\right\}+I_{\rm bg}
	\label{eq:totalscattering}
\end{equation}
where $I_{\rm bg}$ is a constant background, $\left|F\left({\bf \tau}\right)\right|$ the structure factor of the Bragg reflection, $\mathcal{S}_{\rm Bragg}\left({\bf \tau}\right)$ a normalized two-dimensional Gaussian and $c$ is the concentration of defects leading to diffuse scattering modeled by the function $\mathcal{S}_{\rm Diff}\left({\bf Q}\right)$. Introducing the Fourier transform ${\bf s}\left({\bf q}\right)$ of the displacement field ${\bf s}\left({\bf r}\right)$, the diffuse scattering can be modeled by the following function:
\begin{equation}
	\mathcal{S}_{\rm Diff}\left({\bf Q}\right) =
 \left\{\left|{\bf \tau}\cdot{\bf s}\left({\bf q}\right)\right|^{2}-2i{\bf \tau}\cdot{\bf s}\left({\bf q}\right) \, L_{\tau}({\bf q})\right\} \, e^{-2 \, L_{\tau}({\bf q})}
	\label{eq:diffusescattering}
\end{equation}
which is composed of symmetric $\left|{\bf \tau}\cdot{\bf s}\left({\bf q}\right)\right|^{2}$ and antisymmetric $i{\bf \tau}\cdot{\bf s}\left({\bf q}\right)$ parts and involves the static Debye factor of the defects $L_{\tau}({\bf q})$. In the following, we assume that the strain field ${\bf s}\left({\bf r}\right)$ is isotropic around a defect (see \cite{Larson1974,Dederichs1973} along with Supplemental Information). The disorder is parametrized by spherical defects of typical volume $V_{C}$ creating an internal local pressure on the lattice $P_0$. We assume $V_{C} = 10.92~\text{\AA}^{3}$, corresponding to the difference in atomic volume of \pr\,
and \zr\,
ions and $P_{0}$ = 10 eV, which is a typical value in a number of materials \cite{Ehrhart1974,Burkel1979}.

Around each Bragg peak, the diffuse scattering was described by the model given by Eq. (\ref{eq:totalscattering}) using the above fixed parameters $V_{C}$ and $P_0$, and taking the structure factor $F({\bf \tau})$ and the concentration of defects $c$ as fitting parameters. Note that in the course of our analysis, we neglect the effects of sample mosaic spread and imperfect instrumental resolution. Indeed, a two-dimensional Gaussian fit of the nuclear Bragg peaks yields widths much smaller than data pixelation, rendering a convolution of the calculated $I\left({\bf Q}\right)$ with the instrumental resolution function unnecessary (see Supplemental Information).

Examples of fits are given in Fig. \ref{DiffuseNSF} for selected Bragg peaks, together with the calculated part of the Huang diffuse scattering. The fitted structure factor of the Bragg peaks nicely corresponds to calculations of the perfectly ordered structure. The concentration of defects can be estimated to $c = 1.1(5) \cdot 10^{-3}$, when averaged over the 7 measured Bragg peaks. This yields an average volume change induced by one defect of $\Delta V / V \simeq 1.3 \cdot 10^{-3}$, namely an order of magnitude hardly measurable by conventional diffraction (but which could be accessed by neutron Larmor diffraction, which allows measuring the intrinsic width of the unit cell volume distribution \cite{Martin2011,Ruminy2016}).\\

According to Refs. \onlinecite{malkpc,malkin}, the above model corresponds to a random strain field described by the distribution function:
\begin{equation} \label{ge}
g({\bf e}) = \frac{3}{4 \pi^2}\ \frac{\gamma}{(e^2+\gamma^2)^{5/2}},
\end{equation}
where ${\bf e}=\{e_1,e_2,e_3,e_4\}$ denotes the four independent components of the deformation tensor, $e^2 = \sum_{i=1,..,4} e_i^2$ is the total strain and $\gamma$ is a material dependent dimensionless width that reflects the level of defects. Following Malkin {\it et al.} \cite{malkin}, $\gamma$ can be computed from the unit cell volume $V$ dependence on point defect concentration $c$ using the relationship: 
\begin{equation} \label{gamma_Malkin}
\gamma = \frac{\pi\,\left(1+\sigma\right)}{27\,\left(1-\sigma\right)} \cdot \frac{d\ln V}{d\ln c} \quad , 
\end{equation}
where $\sigma$ is the Poisson ratio. Unfortunately, $V(c)$ is unknown for \przr\ so that $\gamma$ cannot be \emph{accurately} determined from the above neutron data only. However, an analysis of existing data on the sister compound \tbti\, \cite{Nakanishi2011} (see Supplemental Information) yields an estimate of the order of magnitude of $\gamma$, namely $10^{-5}-10^{-4}$. Inputing the above refined values for $\Delta V / V$ and $c$ for our \przr\ sample in Eq. \ref{gamma_Malkin}, we indeed get $\gamma \simeq 3.2 \pm 1.5 \cdot 10^{-5}$.

This polarized neutron study thus highlights the existence of strains in the material, not detectable with standard diffraction measurements. With this result in hand, we now examine the consequences of those random strains on the low temperature magnetic properties. 


\subsection{The picture of isolated doublets}

As a first step towards this objective, we consider a simplified picture, where the \pr\ are supposed independent, yet split by a random perturbation due to the strain. Using, at each site, a pseudo spin 1/2 $(\sigma^x,\sigma^y,\sigma^z)$ spanning the $\left\{\left|\uparrow\rangle\right.,\left|\downarrow\rangle\right.\right\}$ subspace, the magneto-elastic interaction takes the simple following form (see Supplemental Information):
\begin{equation}
{\cal H}_{\rm m-el} = \sum_i v_i ~{\sf \sigma}^+_i + v_i^* ~{\sf \sigma}^-_i
\label{h2}
\end{equation}
where $v_i$ is an auxiliary random field that depends on the strain distribution $g({\bf e})$. The real and imaginary parts of the random field $v$ are characterized by an average $\overline{v}=0$ and a standard deviation (which physically corresponds to the strength of the disorder) 
\begin{equation}
\delta v\approx k \gamma
\end{equation}
where $k$ is a numeric constant which merges the magneto-elastic coefficients ($k \sim 1.6 \times 10^4$ K, see Supplemental Information). 

\begin{figure*} [!ht]
\center{\includegraphics[width=\textwidth]{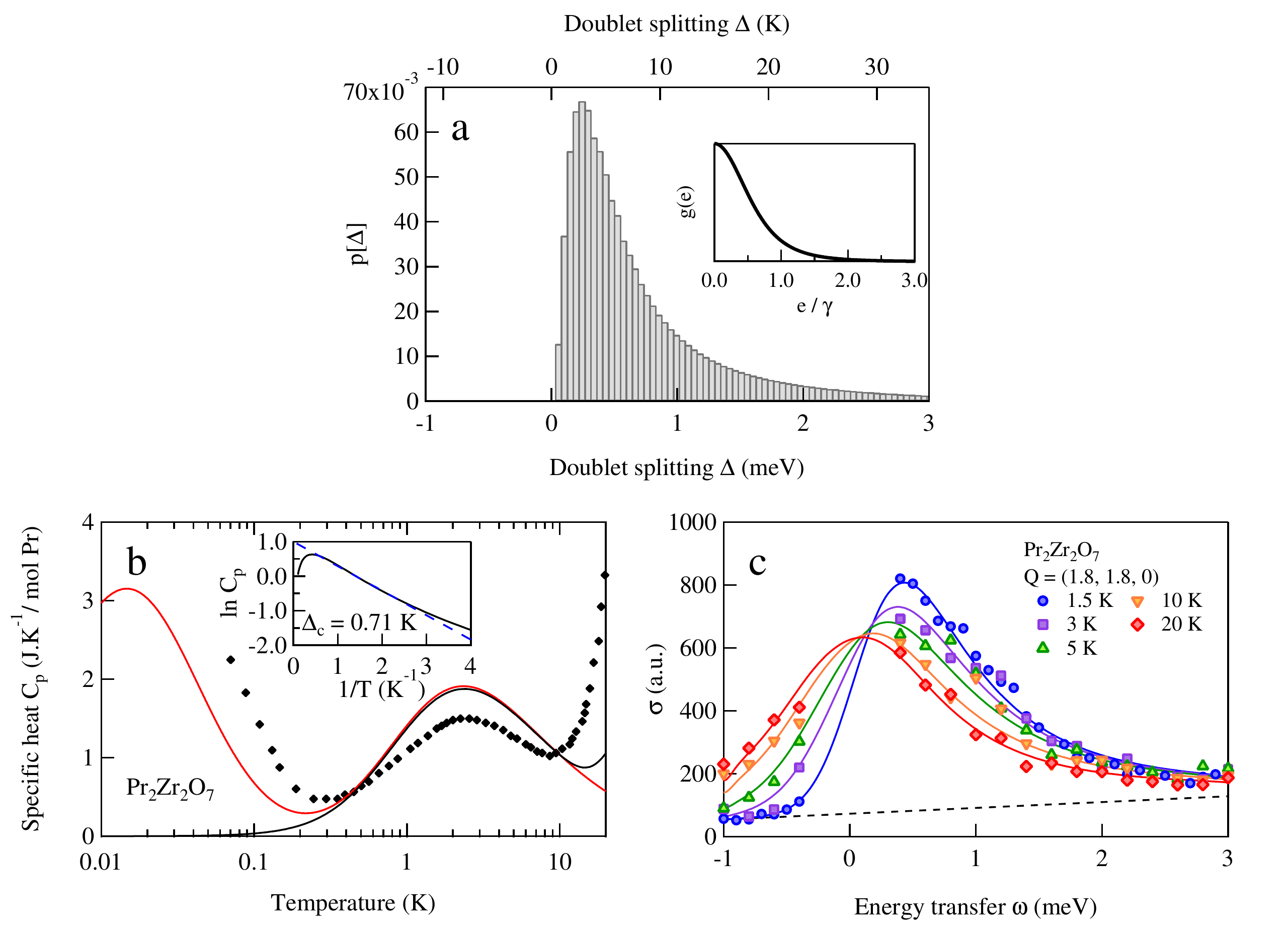}}
\caption{\label{Cp_and_INS} 
{\bf Specific heat and inelastic neutron scattering on \przr.} 
{\bf (a)} Histogram of splitting values $\Delta$ computed using the magneto-elastic Hamiltonian (given in supplemental information) with $B_{21}^{xx}=-$2000\,K, $B_{21}^{zx}=$8000\,K, $B_{22}^{xx} = -$5000\,K, $B_{22}^{zx}=$4000\,K and a width $\gamma=1.25\times 10^{-4}$ of the strain distribution $g(e)$ (inset). {\bf (b)} Calculated specific heat due to a distribution of doublet splittings described by the histogram \textbf{(a)} (black solid line) and experimental data from Ref. \onlinecite{Kimura13}. The activation-like behaviour with an energy $\Delta_c \simeq 0.72$\,K, observed in Ref. \onlinecite{Kimura13} between 0.2 and 2\,K, is correctly reproduced in the simulation (see the blue dashed line in the inset, which is a linear law with slope $\Delta_c$ = 0.71\,K.). The red line shows the specific heat obtained by considering the coupled system of the $^{141}$Pr nuclear spin $I=5/2$ and the split ground electronic doublet. 
{\bf (c)} Neutron scattering cross section measured at different temperatures at {\bf Q}=(1.8,1.8,0). Solid lines are the result of a global fit to the data of the theoretical cross section calculated using the distribution of splittings $\Delta$ shown in {\bf (a)} (see text).
}
\end{figure*}

At this simplified level of approximation, the model given by Hamiltonian (\ref{h2}) describes an assembly of {\it uncoupled} two-levels subsystems. The $| \uparrow \rangle$ and $| \downarrow \rangle$ states recombine in $|a\rangle$ and $|s\rangle$, split by the random variable $\Delta=2|v|$. The shape of the probability density $p[\Delta]$ is shown in Fig. \ref{Cp_and_INS}a. It exhibits an asymmetric profile with a maximum at $\Delta_m \approx 4/3~\delta v \approx 4/3~k \gamma$. 

In the two following paragraphs, the specific heat along with $S({\bf Q},\omega)$, calculated on the basis of Hamiltonian (\ref{h2}) and weighted according to $p[\Delta]$, are compared to experiments. 

\subsubsection{Specific heat}

The specific heat $C_p$ was measured in Ref. \onlinecite{Kimura13} from very low temperature up to 25\,K as well as in Ref. \onlinecite{Petit16} as a function of field. It shows a broad peak between 0.3\,K and 10\,K, followed at very low temperature by a steep upturn attributed to hyperfine effects due to the nuclear \pr\ spin. 

In the above model of independent split doublets, the specific heat is written:
\begin{equation}
C_p = k_{\rm B} \sum_{\Delta} p\left[\Delta\right] (\frac{\Delta}{2k_{\rm B} T})^2 \frac{1}{\cosh^2 \frac{\Delta}{2k_{\rm B} T}}
\end{equation}
Our simulation of the specific heat $C_p(T)$ is shown in Fig. \ref{Cp_and_INS}b (black solid line, together with the $1/T$ plot of $\ln C_p(T)$ in inset). It consists of multiple Schottky peaks due to the superposition of individual random splittings $\Delta$ induced by local strains. This contrasts with the interpretation suggested in Ref. \onlinecite{Kimura13} in terms of monopole quantum dynamics. We find that for a value $\gamma$ = 1.25 $\times 10^{-4}$ (hence within the uncertainty range stimated from Huang scattering), corresponding to a disorder strength $\delta v \approx$ 2\,K, $C_p(T)$ shows a peak at 2\,K, matching the experimental result, with a ($\simeq$ 25 \%) overestimate of the peak amplitude. The entropy released by the calculated specific heat anomaly is expectedly R$\ln 2$, which is larger than the value estimated from the $4f$ contribution extracted from the data \cite{Kimura13}, and in agreement with the higher calculated peak value. 

In Ref. \onlinecite{Kimura13}, it was proposed that the upturn observed at lower temperature could be tentatively attributed to a hyperfine Schottky anomaly. However, this work pointed out that, provided the ground \pr\ doublet remains degenerate, this hypothesis would lead to a too large upturn, the specific heat showing then a maximum at 0.14\,K peaking at 7\,JK$^{-1}$(mol.Pr)$^{-1}$. We show here that the disorder not only allows to qualitatively explain the peak at 2 K, but also resolves this discrepancy at very low temperature. The simulation is performed by adding the hyperfine contribution $A \ {\bf I}_z \ {\bf J}_z $ to the Hamiltonian, where $A$ is the magnetic hyperfine constant $A \simeq$0.055\,K \cite{bleaney} and ${\bf I}_z$ is the Pr nuclear spin with quantum number $m = 5/2, 3/2, \dots ,-5/2$. The calculated specific heat is represented as a red curve in Fig. \ref{Cp_and_INS}b, revealing that the hyperfine upturn has the correct order of magnitude. It is worth mentioning that the magnitude of this upturn directly results from the distribution of splittings: when the strain is small, the ground \pr\ doublet survives and remains degenerate, with an Ising behavior. This gives rise to a large hyperfine anomaly of the specific heat. In contrast, when the strain is large, the ground state is a singlet and the hyperfine anomaly weakens. The existence of a splitting distribution thus creates an intermediate situation between these two limiting cases, and finally explains why the upturn of the specific heat at low temperature remains moderate. In this sense, the fact that the upturn remains somewhat underestimated indicates that $\delta v \approx$ 2\,K is likely an upper limit of the actual strain distribution. We also notice that the amplitude of the low temperature up-turn strongly depends on the sample synthesis, namely the growth rate of the crystal \cite{Koohpayeh2014}.

\subsubsection{Spin excitations and Inelastic neutron scattering}

Low energy inelastic neutron scattering spectra taken from Refs. \onlinecite{Guitteny15,Petit16}, are represented in Fig. \ref{Cp_and_INS}c at selected temperatures and for a momentum transfer {\bf Q}=(1.8,1.8,0). They are part of the dynamical spin ice mode reported initially in Ref. \onlinecite{Kimura13} and consist of a broad inelastic line whose width increases and whose peak intensity decreases upon heating. Both quantities saturate above $\simeq$ 10\,K up to 50\,K. 

In the model of independent doublets defined by Eq. (\ref{h2}) and $p[\Delta]$, $S({\bf Q},\omega)$ is given by (see Supplemental Information):
\begin{eqnarray} \label{cross}
S({\bf Q},\omega) & \simeq & (1+n(\omega)) \, \zeta^2 \times \nonumber\\
& &\sum_{\Delta} p\left[\Delta\right] \tanh(\frac{\Delta}{2k_{\rm B}T})\ G(\Delta,\Gamma) 
\end{eqnarray}
where $1+n(\omega) = [1-\exp(-\omega/k_{\rm B}T)]^{-1}$ is the detailed balance factor and $G(\Delta,\Gamma)$ represents energy profiles characterized by differences of Lorentzian line-shapes centered at $\pm \Delta$ with a half width at half maximum (HWHM) $\Gamma$. $S({\bf Q},\omega)$ thus consists of a series of modes at the random energies $\Delta$ and the global width of the signal is then a direct signature of the randomness. $\zeta$ is defined as $|\langle a | {\bf J} | s \rangle|^2$ and owing to the actual CEF, $\zeta =$ 3.4. The neutron cross section is finally given by 
\begin{equation} \label{cross1}
	\sigma = a \cdot S({\bf Q},\omega) \ast R({\bf Q},\omega) + {\rm bg}(\omega)
\end{equation}
where $a$ is an instrumental scaling factor, ${\rm bg}(\omega)$ a linear background term and $R({\bf Q},\omega)$ the TAS resolution function (see Supplemental Information). It should be stressed here that the non zero cross section is due to the recombination of the $\uparrow$ and $\downarrow$ states. Otherwise, it would be zero because of the non-Kramers nature of the \pr\, ion which imposes $\langle \uparrow | {\bf J} | \downarrow \rangle \equiv 0$.

The calculated cross sections at different temperatures are represented in Fig. \ref{Cp_and_INS}c with the \emph{same} $p[\Delta]$ as the one used to calculate the specific heat. As expected from an ensemble of non-degenerate doublets, the overall scattering is {\it inelastic}, with a peak around 0.4\,meV at the lowest temperatures. Comparing to the experimental data, the spectra at 1.5, 3, 5, 10 and 20\,K are indeed satisfactorily reproduced by a simultaneous fit of Eq. \ref{cross1} to the whole set of curves. Since the physics probed by inelastic neutron scattering is assumed to be driven by random strains, hence by $p[\Delta]$, it is actually natural for $a$ to be $T-$independent. This agreement implies that the model captures the physics at play. In turn, the thermal increase of the line width is thus due to scattering by a distribution of electronic transitions with energies being of the same magnitude as $T$.

The above results advocate that the picture of independent doublets split by the \emph{actual} disorder can account for a number of features not understood so far: provided that the strength of the disorder $\delta v$ is of the order of 1 K (with a typical strain $\gamma \approx 1.25 \times 10^{-4}$), the temperature dependence of the specific heat is captured as well as the broadening of $S({\bf Q},\omega)$. Note that this value of $\delta v$ should be considered as an estimate only: indeed, on the one hand, $\gamma$ likely depends on the sample preparation; on the other hand, $\delta v$ also depends on the exact values of the magneto-elastic coefficients via the coefficient $k$ (see Supplemental information). Furthermore, the exact profile of $p[\Delta]$ may be more complex. We note for instance that the distribution of splitting proposed by Wen \emph{et al} \cite{Wen2017} on the basis of INS data, is similar to $p[\Delta]$, yet reaches a maximum at $\Delta=0$. It would be interesting to see how this affects $C_p$ for instance. 

It remains that as far as the doublets are independent, no particular feature is expected in reciprocal space, so that $S({\bf Q},\omega)$ should not be ${\bf Q}$ dependent in this approach. Experimentally, though, $S({\bf Q},\omega)$ displays an inelastic spin-ice like pattern and is thus ${\bf Q}$-dependent. A thorough description of the actual ground state of \przr\, should then incorporate both the randomness and interactions between the pseudo-spins. 

\subsection{The picture of coupled doublets}

\begin{figure*}
\center{
\includegraphics[width=\textwidth]{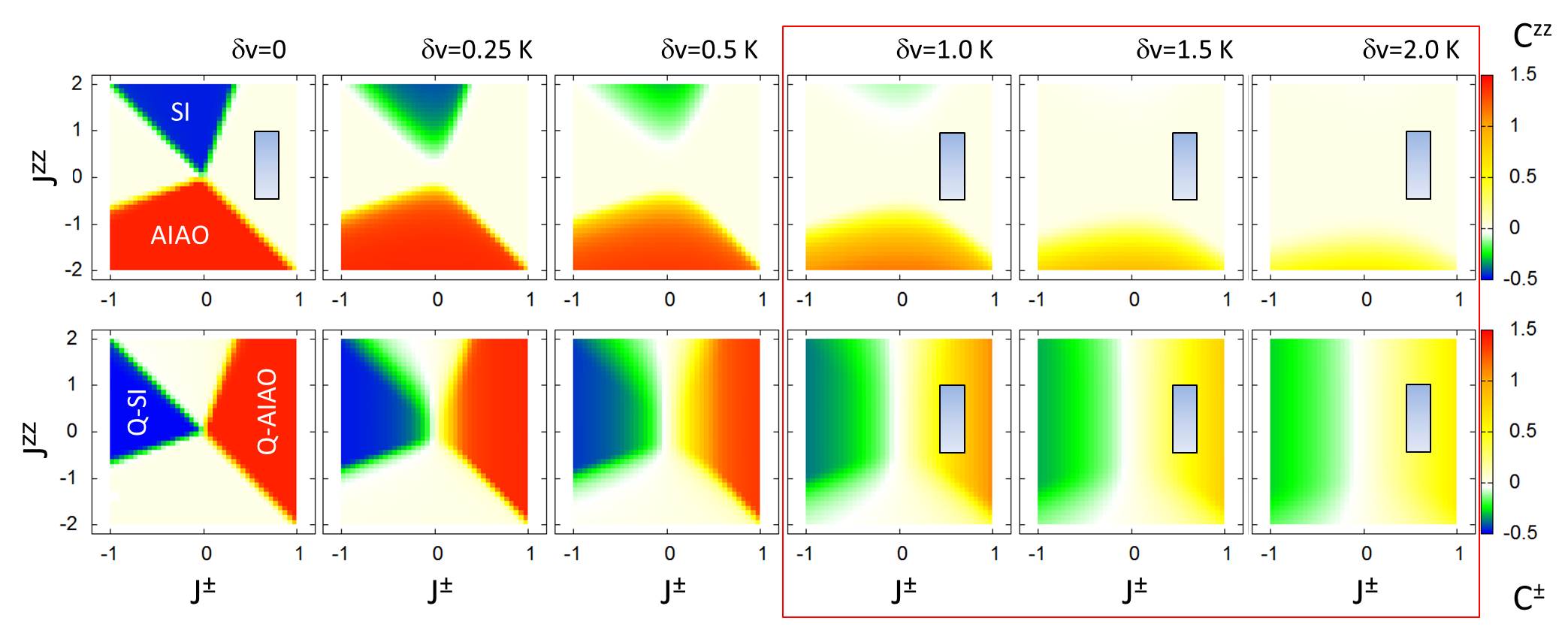}
}
\caption{\label{disorder} {\bf Influence of disorder on the mean field phase diagram}. The upper and lower panels display the magnetic and quadrupolar correlation functions $C^{zz}$ and $C^{\pm}$ as a function of ${\cal J}^{\pm,z}$ and for different disorder strength $\delta v$. Without disorder ($\delta v=0$), these quantities provide a picture of the mean field phase diagram. This phase diagram encompasses two dipolar phases, spin ice (SI) and ``all$-$in - all$-$out'' (AIAO) as well as two quadrupolar ordered phases, Q-SI and Q-AIAO. In the latter, the pseudo spin is perpendicular to the ${\bf z}_i$ axis. The rectangle displays the region of the phase diagram proposed for \przr\, in Ref. \onlinecite{Petit16}. According to our definitions, the $\sigma^z$ are of the same sign in the AIAO states and staggered in the SI phase. $C^{zz}$ then reaches its maximum in the long range ordered phase with $C^{zz}=1/2 \times 6\times 1/2 = 3/2$ and $C^{zz}= 1/2 \times (-4+2) \times 1/2= -1/2$ for the AIAO and SI phases respectively. Similarly the $\sigma^x$ have the same sign in the Q-AIAO states and are staggered in the Q-SI phase. Again, the maximum value is reached in the long range ordered phase with $C^{\pm}=3/2$ and $C^{\pm}= -1/2$ respectively. The gray rectangles show the range of parameters proposed for \przr. The red rectangle emphasizes the results for the value of $\delta v$ estimated from specific heat and neuton scattering. 
}
\end{figure*}

To go in this direction, we follow Ref. \onlinecite{Onoda10,Onoda11,Lee12} to take couplings into account. The magnetic and quadrupolar degrees of freedom are described in a unified framework by considering the components $(\sigma^x,\sigma^y,\sigma^z)$ of the pseudo spins 1/2 that span the $\left\{\left|\uparrow\rangle\right.,\left|\downarrow\rangle\right.\right\}$ subspace. Provided these states are well protected from the first excited state, the Ising magnetic moments ${\bf J}_{z,i}$, pointing along the $\langle 111 \rangle$ ${\bf z}_i$ axes, as well as the quadrupoles operators ${\bf Q}_{x^2-y^2} = {\bf J}_x^2 - {\bf J}_y^2$, ${\bf Q}_{xy} = {\bf J}_x {\bf J}_y + {\bf J}_y {\bf J}_x$, ${\bf Q}_{xz} = ({\bf J}_x {\bf J}_z + {\bf J}_z {\bf J}_x)/2$ and ${\bf Q}_{yz} = ({\bf J}_y {\bf J}_z + {\bf J}_z {\bf J}_y)/2$ can be mapped onto the $(\sigma^x,\sigma^y,\sigma^z)$ Pauli matrices (see Supplemental Information) via the $q_{\parallel,\perp}$ coefficients:
\begin{eqnarray}
{\bf J}_z & = & 2\zeta ~\sigma^z ~\mbox{with}~2\zeta=\frac{g_{\parallel}}{g_{\rm J}} \nonumber \\
{\bf Q}_{x^2-y^2} & = & {\bf J}_x^2 - {\bf J}_y^2 = q_{\parallel} \sigma^x \nonumber \\
{\bf Q}_{xy} & = & {\bf J}_x {\bf J}_y + {\bf J}_y {\bf J}_x = q_{\parallel} \sigma^y \nonumber \\
{\bf Q}_{xz} & = & ({\bf J}_x {\bf J}_z + {\bf J}_z {\bf J}_x)/2 = q_{\perp} \sigma^x \nonumber \\
{\bf Q}_{yz} & = & ({\bf J}_y {\bf J}_z + {\bf J}_z {\bf J}_y)/2 = q_{\perp} \sigma^y \nonumber 
\end{eqnarray} 
$g_J=4/5$, $g_{\parallel}$ is an effective anisotropic $g$ factor, and, together with $q_{\parallel,\perp}$, depend on the actual CEF scheme ($g_{\parallel}/g_{\rm J} \approx$ 6.8, $\zeta \approx 3.4$, $q_\perp \approx$ 0.055 and $q_{\parallel}\approx$ 2.17). On this basis, a bilinear Hamiltonian has been proposed \cite{Rossprx11,Savary12,Curnoe07,Curnoe14,Onoda10,Onoda11,Lee12}: 
\begin{eqnarray}
{\cal H} &= &\frac{1}{2} \sum_{<i,j>} {\cal J}^{zz} {\sf \sigma}^z_{i} {\sf \sigma}^z_{j} 
+ \frac{1}{2} \sum_{<i,j>} -{\cal J}^{\pm} \left(\sigma^+_i \sigma^-_j + \sigma^-_i \sigma^+_j\right) \nonumber \\
&+&\frac{1}{2} \sum_{<i,j>} {\cal J}^{\pm\pm} \left(\gamma_{ij} \sigma^+_i \sigma^+_j + \gamma^*_{ij}\sigma^-_i \sigma^-_j \right) 
\label{h}
\end{eqnarray}
The $\gamma_{ij}$ parameters are unimodular matrix coefficients defined in Ref. \citenum{Rossprx11}. ${\cal J}^{zz}$, ${\cal J}^{\pm}$ and ${\cal J}^{\pm\pm}$ are the effective dipolar and quadrupolar exchange terms, compatible with the local symmetry of the rare earth. Meanwhile, estimates of ${\cal J}^{\pm}$ and ${\cal J}^{zz}$ have been proposed in Ref \onlinecite{Petit16}:
\begin{eqnarray*}
0.7 \leq {\cal J}^{\pm} \leq 0.8 \textrm{ K} \\
-0.5 \leq {\cal J}^{zz} \leq 1 \textrm{ K}
\end{eqnarray*}
$\delta v$, ${\cal J}^{\pm}$ and ${\cal J}^{zz}$ are then of the same order of magnitude, reinforcing the idea that the picture of {\it decoupled} doublets is indeed simplistic. The full Hamiltonian ${\cal H}+{\cal H}_{\rm m-el}$ (Eq (\ref{h}) and Eq. (\ref{h2})) then describes an anisotropic Heisenberg model in a random transverse field $v_i$ whose distribution is ruled by $\gamma$. 

The exact solution to this problem is beyond the scope of the present work. We shall however carry on with a simplified picture and consider the mean field approximation (assuming ${\cal J}^{\pm\pm}=0$ as in Ref \cite{Petit16}):
\begin{equation}
\label{mf}
{\cal H}+{\cal H}_{\rm m-el} \approx \sum_i V_i ~{\sf \sigma}^+_i + V_i^* ~{\sf \sigma}^-_i + \eta_i ~{\sf \sigma}^z_i 
\end{equation}
with $V_i= v_i - \sum_j {\cal{J}}^{\pm} \langle {\sf \sigma}^-_j \rangle $ and $\eta_i = \sum_j {\cal{J}}^{zz} \langle {\sf \sigma}^z_j \rangle$. At this level of approximation, the model describes an assembly of {\it coupled} two-levels subsystems. The eigenstates $|\pm \rangle_i$ (at a given site $i$) are split by the (positive) random quantity $\Delta_i = 2\sqrt{|V_i|^2+\eta_i^2/4}$ and can be written in a convenient way using local spherical angles $\phi_i$ and $\theta_i$:
\begin{eqnarray*}
|+ \rangle_i &=& e^{i\phi_i} \sin \frac{\theta_i}{2} | \downarrow \rangle_i + \cos \frac{\theta_i}{2} | \uparrow \rangle_i \\
|- \rangle_i &=& -e^{-i\phi_i} \sin \frac{\theta_i}{2} | \uparrow \rangle_i + \cos \frac{\theta_i}{2} | \downarrow \rangle_i
\end{eqnarray*}
The $|\pm \rangle_i$ states can be seen as intermediate states between the ``tunnel-like'' wave-functions $\vert s,a \rangle$ given by Eq. \ref{tunn} and the original $\vert \uparrow \downarrow \rangle$ doublet. $\phi_i$ and $\theta_i$ have a natural physical interpretation: $\pm\frac{1}{2}\cos \theta_i$ is the dipolar magnetic moment at site $i$ and $\pm\frac{1}{2}\sin \theta_i ( \cos \phi_i , \sin \phi_i)$ is the ``quadrupolar'' moment at the same site. Their values as well as whether $|+\rangle$ or $|- \rangle$ is the ground state depend on the parameters ${\cal J}^{\pm}$, ${\cal J}^{zz}$ and on the $v_i$ random field. 

In the absence of disorder, the mean field phase diagram consists of long range ordered magnetic and quadrupolar phases (see Ref \onlinecite{Petit16} along with the left column of Figure \ref{disorder}). It encompasses an antiferromagnetic ``all$-$in - all$-$out'' phase (AIAO) for ${\cal J}^{zz} \le 0$ (the four spins of a tetrahedron pointing out or towards the center), an ordered spin ice phase (SI) for ${\cal J}^{zz} \ge 0$ (two spins pointing out and two spins towards the center), and two quadrupolar phases denoted Q-SI for ${\cal J}^{\pm} \le 0$ and Q-AIAO for ${\cal J}^{\pm} \ge 0$. In the magnetic phases, the pseudo spin are aligned along the local ${\bf z}_i$ axes ($\theta = 0, \pi$). In contrast, in the quadrupolar phases, the pseudo spin 1/2 are tilted away and order within the XY plane ($\theta = \pi/2$). 

On top of this fully ordered background, the pseudo spins tend to get additional random $\sigma^x$ and $\sigma^y$ components owing to strain. Clearly, the net ordered moment along the ${\bf z}_i$ axes or within the XY plane will decrease with increasing disorder and may even be averaged to zero for strong disorder. Depending on the values of ${\cal J}^{zz}$ and ${\cal J}^{\pm}$, magnetic (dipolar) or quadrupolar correlations may or may not survive. It is then convenient to discuss the phase diagram using the magnetic and quadrupolar correlations functions:
\begin{eqnarray*}
C^{zz} & = & \frac{1}{N} \sum_{i,\langle j \rangle_i} \langle \sigma^z_i \rangle \langle \sigma^z_j \rangle \\
C^{\pm} &=& \frac{1}{N} \sum_{i,\langle j \rangle_i} \langle \sigma^+_i \rangle \langle \sigma^-_j \rangle
\end{eqnarray*}
the sum over ``$\langle j \rangle_i$'' corresponds to a sum over the neighbors connected by ${\cal{J}}^{zz}$ or ${\cal{J}}^{\pm}$. Figure \ref{disorder} shows $C^{zz}$ and $C^{\pm}$ as a function of ${\cal J}^{\pm}$ and ${\cal J}^{zz}$ for different disorder strength. Positive (resp. negative) values of $C^{zz}$ indicate AIAO (resp. SI) correlations. In a similar way, positive (resp. negative) values of $C^{\pm}$ indicate Q-AIAO (resp. Q-SI) correlations. With increasing disorder, the locations of the different long range ordered phases are still visible but all the more smeared out that ${\cal{J}}^{zz} \ll \delta v$ or ${\cal{J}}^{\pm} \ll \delta v$. In the latter regions of the phase diagram, the long range ordered phases are unstable, being replaced by disordered phases where short range correlations survive (note that the AIAO correlations have a better resistance to disorder than the SI ones essentially because the molecular field is larger in the former). 

\begin{figure*}
\center{
\includegraphics[width=\textwidth]{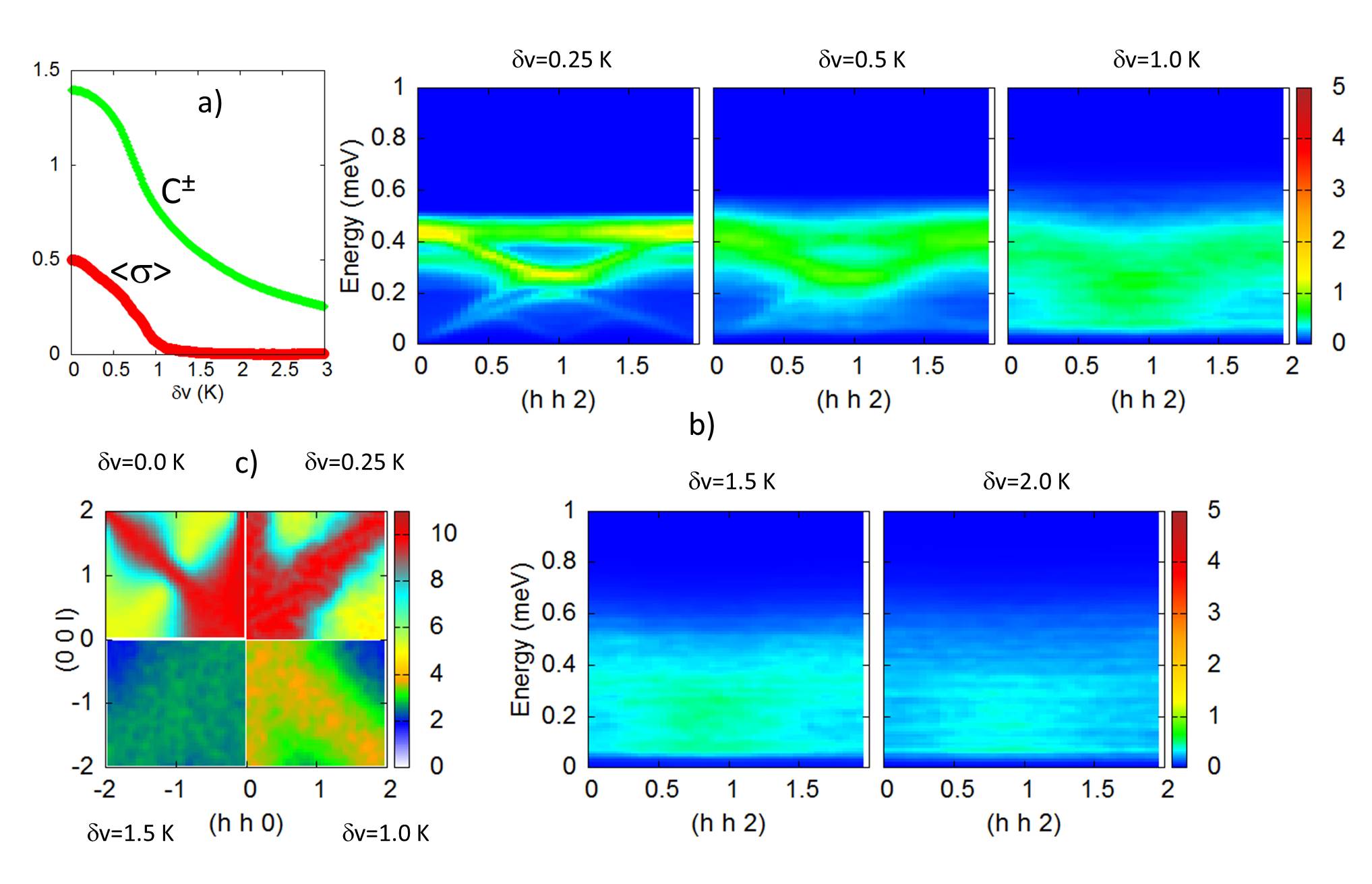}
}
\caption{\label{spindyn} {\bf Influence of disorder on the spin dynamics}. {\bf (a)} shows the disorder dependence of the Q-AIAO order parameter and of the $C^{\pm}$ correlation function. {\bf (b)} shows $S({\bf Q},\omega)$ along $Q=(h, h, 2)$ for different disorder strengths, ranging from $\delta v= 0.25$ up to $\delta v= 2$ K, i.e. in absence of magnetic or quadrupolar orderings. {\bf (c)} displays $S({\bf Q},\omega)$, integrated in the range $0.38\leq \omega\leq 0.48$ meV, and presented in the $(hh0)-(00,\ell)$ reciprocal lattice plane. The arm-like features are well identified for the different disorder strength altough weaker and less contrasted with increasing $\delta v$. The upper left panel in {\bf c}) shows the dynamical spin ice pattern for $\delta v=0$. The color scale is chosen to observe the arms 
more easily.}
\end{figure*}

For $\delta v = 2$ K and the coupling parameters proposed for \przr\, (${\cal J}^{\pm}= 0.7$ K and ${\cal J}^{zz}= -0.5$ K), the Q-AIAO order parameter is already suppressed (see Figure \ref{spindyn}a), but antiferroquadrupolar correlations on neighboring sites are found to survive (see Figure \ref{disorder}). However, owing to uncertainties on the precise value of $\delta v$, we should not confine ourselves to $\delta v = 2$ K: from the phase diagram considered as a whole, it can be inferred that \przr\, locates in a region where the ground state is disordered or just about to order in the long range Q-AIAO state. Calculations of $S({\bf Q},\omega)$ based on real space time evolving spin dynamics simulations (see methods) \cite{Robert2008} have been carried out for different $\delta v$. For $\delta v = 0$, the calculations reproduce the dynamical spin-ice mode observed at 0.4 meV initially reported in Ref. \onlinecite{Kimura13}, with its typical arm-like features in reciprocal space. The spectra along $(h, h, 2)$ shown in Figure \ref{spindyn}b illustrate that with increasing disorder, the energy width of the response grows rapidly, resulting in spectra similar to the ones shown in Figure \ref{Cp_and_INS}c once convolved with the experimental resolution. The dispersive features also become difficult to distinguish. By integrating $S({\bf Q},\omega)$ within $0.38\leq \omega\leq 0.48$ meV, we obtain the $Q$-dependence of the dynamical spin ice mode show in Figure \ref{spindyn}c. The important point is that the arm-like features, although weaker and marked downturn, are still visible for strong disorder, i.e. in the disordered phase. The pinch points are also considerably blurred, which is consistent with the experiments.


\section{Discussion and Summary}

As detailed in this paper, we thus propose that the randomness in {\it non Kramers ions} based pyrochlore magnets (such as \pr\ and \tb), promotes the rise of new disordered phases, characterized by short range magnetic or quadrupolar correlations. More precisely, by virtue of the magneto-elastic coupling, strains split the non-Kramers doublet of the rare earth ion in a random fashion, recombining the $\uparrow$ and $\downarrow$ Ising wavefunctions in ``tunnel-like" states. ${\cal J}^{\pm}$ and $ {\cal J}^{zz}$ compete with those disorder effects, fostering correlations among dipoles and/or among quadrupoles, deep in the disordered phases, and stabilizing magnetic and/or quadrupolar orderings, at the mean field level, for weak disorder.

Our estimation of the disorder strength $\delta v$ in \przr, based on the analysis of polarized diffuse scattering maps and on the fit of various experiments, gives a value of the same order of magnitude as ${\cal J}^{\pm}$ and $ {\cal J}^{zz}$, locating this material in an intermediate to strong coupling regime.

This scenario provides a very likely and qualitative explanation for a number of experimental features reported in \przr, such as the lack of long range order, the temperature dependence of the specific heat (including the peak at 2K as well as the upturn at lower temperatures), the energy broadening of the spin excitations spectrum observed by neutron scattering and its specific spin-ice like ${\bf Q}$-dependence. 

It is worth noting that a related model has been proposed in Ref \onlinecite{Savary2017}, with ${\cal J}^{\pm}\equiv 0$ though, hence neglecting quadrupolar effects, the disorder being modeled by a transverse random field $h$, with average $\bar{h}$ and standard deviation $\delta h$. Based on a sophisticated treatment, this study points out that disorder provokes quantum superpositions of spins throughout the system, entailing the rise of an emergent gauge structure along with a set of fractional excitations. It also predicts the existence of a ``Coulombic Griffiths phase'' for large $\delta h$ and $\bar{h}\ne 0$. Interestingly, this case resembles, to some extent, to the ${\cal J}^{\pm}\ne 0$ case considered in the present work. Indeed, at the mean field level, in the Q-AIAO phase, $V_i=v_i-\sum_j{\cal J}^{\pm}\langle \sigma_j^{-}\rangle$ behaves as the random field $h$ with a non-zero $\bar{h}$ average. As a result, even if dedicated theoretical developments are necessary for a rigorous justification, \przr\, could be a relevant candidate for this new state of matter called ``Coulombic Griffiths phase''.

A few experimental facts remain, however, open questions. For instance, the low temperature upturn of the magnetic susceptibility \cite{Petit16,Sibille16} as well as the, although extremely weak, spin-ice elastic response \cite{Kimura13} are not captured, at least with these coupling parameters. We emphasize, however, that a positive ${\cal J}^{zz}$ could help understanding these issues, since such a value would lead to correlated magnetic moments in the spin-ice manner. Additional experiments are required to shed light on this point. Furthermore, it would be fruitful to consider other forms of disorder, as for instance the bond disorder pointed out in a recent study of \tbhf \cite{Sibille16}.

As it opens an original research route in the field of quantum spin ice and spin liquids, we hope that this work will arouse new theoretical developments taking especially into account the role of ${\cal J}^{\pm}$. We also anticipate that experiments on other non-Kramers quantum spin ice candidates as \prsn\,\cite{Princep13}, \prhf \cite{Sibille2017} and \tbhf, where a strongly fluctuating Coulomb liquid phase coexists with defect induced frozen magnetic degrees of freedom \cite{Sibille16}, will be extremely interesting to test these ideas. 

\section*{Acknowledgements}

Authors gratefully acknowledge ILL and LLB for beam-time allowance. We also thank B. Z. Malkin for his help with the magneto-elastic calculations and acknowledge C. Diop for his advices in calculating the distribution functions. The work at the University of Warwick was supported by the EPSRC, UK, through Grant EP/M028771/1.



\bibliography{PrZr}

\end{document}